**On mechanism of high-temperature superconductivity in hydrogen sulfide at 200 GPa: transition into superconducting anti-adiabatic state in coupling to H-vibrations**

P. Baňacký
Chemical Physics division, Institute of Chemistry, Faculty of Natural Sciences, Comenius University, Mlynská dolina CH2, 84215 Bratislava, Slovakia

**Abstract** We have shown that adiabatic electronic structure of superconducting phase of sulphur hydride at 200 GPa is unstable toward vibration motion of H-atoms. Theoretical study indicates that in coupling to H-vibrations, system undergoes transition from adiabatic into stabilised anti-adiabatic multi-gap superconducting state at temperature which can reach 203 K.

**Introduction** Guided by the idea of Ashcroft [1] on metallization of hydrogen and electron-phonon (EP) interaction background of the BCS theory of superconductivity, Drozdov et al. have metalized hydrogen sulfide and at 200 GPa detected superconductivity [2] with the highest critical temperature recorded so far, $T_c \sim 203$ K. Within the Allen-Dynes' strong-coupling treatment, superconductivity of $H_2S$ with such a high $T_c$ was calculated by Duan et al. [3] for stable phase at 200 GPa, which has been theoretically predicted as a bcc crystal structure of Im3-m symmetry with $(H_3S)_2$ composition of unit cell. This structure was recently confirmed experimentally by synchrotron XRD [4]. Several theoretical studies on superconductivity of this phase have been published [5-9] recently. It should be noted that superconductivity aspects are studied within the adiabatic BCS, or Eliashberg modified theories with EP interactions representing crucial mechanism of superconducting state transition. No matter if particular authors emphasize the role of anharmonicity of H-vibrations, Lifshitz transition and zero-point lattice fluctuation along with peculiar electronic structure and character of total density of states (DOS) calculated for equilibrium structure in frozen nuclear configuration, the nonadiabatic effects are expected to be crucial in understanding mechanism of superconductivity in this system. The reason is the fact that electronic and phonon spectra are on comparable energy scales. In particular, for equilibrium geometry at 200 GPa, adiabatic ratio of this system is $\omega/E_F < 1$ (but not $<< 1$) and application of current adiabatic theories of superconductivity is in these circumstances rather problematic.

Present study show that physics of this system is even more complicated when electronic structure is investigated in distorted geometries with H–atoms displacements in S-H stretching and/or bending vibrations. In this case, vibration motion induces instability of adiabatic electronic structure. It can be identified as splitting of degeneracy of bands above Fermi level (FL) connected with fluctuation of one of the band near FL in Γ point. At some vibration displacement, the top of one of bands crosses the Fermi level and in vibration motion when the band-top approaches FL it represents decrease of Fermi energy ($E_F$), resulting in switching into anti-adiabatic regime with $\omega/E_F > 1$. In this case, not only Eliashberg approximation does not hold, but Born-Oppenheimer approximation (BOA) is broken. Theory of electron-vibration interaction beyond the BOA was elaborated in [10] and it has been shown [11-16] for different types of low and high-$T_c$ superconductors (e.g. A15 structures, cuprates and $MgB_2$) that transition into superconducting state is directly related to EP coupling-induced transition from metal-like adiabatic state into stabilized and geometrically degenerated anti-adiabatic ground state connected with gap(s) opening in single-particle spectrum.

**Results** In modeling H-vibrations, effective is to study $(H_3S)_2$ in cP lattice – Figure 1a, with following atom coordinates and displacements δ in unit cell: S1(0,0,0); S2(1/2,1/2,1/2); H1(1/2,0+δ,0); H2(0,1/2-δ,0); H3(0,0,1/2); H4(1/2-δ,0,1/2); H5(0+δ,1/2,1/2); H6(1/2, 1/2,0). For equilibrium structure (δ=0) with lattice parameter a=2.984 Å and frozen nuclear configuration, the electronic band-dispersions along direction Γ[0,0,0] – X[0,1/2,0] is in Figure 1b. Three bands are degenerated above FL in Γ point and Fermi energy is $E_F \approx 0.9$ eV. Calculations of phonon dispersions [3,5,7] yield two distinct sets of phonon branches with H-dominated vibrations; $\omega_H \approx 200$ meV and 150 meV in Γ point. It means that in equilibrium geometry with frozen nuclear positions, adiabatic ratio is $\omega_H/E_F \approx 2/9$, which is basically on the edge of validity of adiabatic approximation. For H-atoms displacements δ=0.02/H, degeneracy of bands is lifted and top of fluctuating band is shifted below FL-Figure 1c. In vibration motion, for δ=0.017/H (i.e. 0.035 Å/H) top of this band is

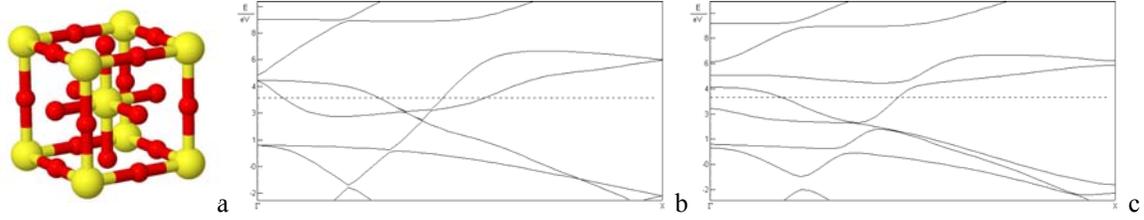

**Figure 1** (a) Crystallographic cell of cubic H$_3$S at 200GPa with two formula unit/u.c. Electronic band structure in (b) equilibrium geometry and (c) in geometry with H-vibration distortions δ=0.02/H – see text still above FL but, energy distance to FL is only 0.123eV (i.e. E$_F$). It indicates formation of anti-adiabatic regime with ω$_H$/E$_F$≈1.6, or 1.2 depending on particular ω$_H$. According to nonadiabatic theory of electron-vibration interactions [10-12], whether system can be stabilized in anti-adiabatic state at distorted geometry depends on correction $\Delta E^0_{(na)}$ to the ground state energy,

$$\Delta E^0_{(na)} = 2\sum_{\varphi_{Rk}}^{bands}\sum_{\varphi_{Sk}}^{bands}\int_0^{\varepsilon^0_{k',max}} n_{\varepsilon^0_{k'}}(1-f_{\varepsilon^0_{k'}})d\varepsilon^0_{k'} \int_{\varepsilon^0_{k',min}}^{\varepsilon^0_{k,max}}|u^r_{k-k'}|^2 n_{\varepsilon^0_k} f_{\varepsilon^0_k} \frac{\hbar\omega_r}{(\varepsilon^0_k - \varepsilon^0_{k'})^2 - (\hbar\omega_r)^2} d\varepsilon^0_k \quad , \quad k<k_F ; k'>k_F \qquad (1)$$

For band structure calculation we have used Cyclic Cluster HF-SCF method [17], and for δ=0.017/H strong electron-vibration coupling (mean value of matrix element) $u^r_{k-k'}$ ≈3.9 eV has been calculated. Partial DOS of fluctuating band when approaching FL is relatively small, $n_{\varepsilon^0_{k'}}$ (FL)≈0.123 and P-DOS of bands inter-sectting FL are even smaller, 0.05. Nonetheless, due to strong el-vibration coupling, $\Delta E^0_{(na)}$ is ≈ -1eV/u.c.

Since increase in total energy due to displacement δ=0.017 on adiabatic level is $\Delta E_d$ ≈+0.1 eV/u.c., the final effect is stabilization of anti-adiabatic state in distorted geometry with fluxionally degenerated ground state structure on lattice scale. Effect on single particle spectrum is also very strong – gap in original metal-like bands at FL are opened. Quasi-degenerate states above FL are shifted upwards,

$$\Delta\varepsilon(Pk') = \sum_{Rk'_1>k_F}|u^q|^2 (1-f_{\varepsilon^0_{k'_1}}) \frac{\hbar\omega^q_{k'-k'_1}}{(\varepsilon^0_{k'} - \varepsilon^0_{k'_1})^2 - (\hbar\omega^r_{k'-k'_1})^2} - \sum_{Sk<k_F}|u^{k-k'}|^2 f_{\varepsilon^0_k} \frac{\hbar\omega^q_{k'-k'_1}}{(\varepsilon^0_{k'} - \varepsilon^0_k)^2 - (\hbar\omega^q_{k'-k'_1})^2} \qquad (2)$$

for $k' > k_F$, and occupied states below FL are shifted in opposite direction-downwards from FL. The gap opened in this way obeys following T-dependence, $\Delta(T) = \Delta(0)tgh(\Delta(T)/4k_BT)$, and for critical temperature holds, $T_c = \Delta(0)/4k_B$. Calculated parameters for ω$_H$≈150 meV are $\Delta(0)$≈62.2 meV and $T_c$≈170 K. For ω$_H$≈200 meV, the value of opened gap is $\Delta(0)$≈74.8 meV and $T_c$≈203 K. It should be stressed that in Γ – X direction 2 gaps are opened at k-points where particular bands intersects FL on adiabatic level- Figure 1c.

**Acknowledgements** This work was supported by the Slovak Research and Development Agency under the contract No. APVV-0201-11.